%% file: jaspap.tex
\documentclass[preprint,11pt]{elsarticle}

\usepackage{epsfig}
\usepackage{array,tabularx,epsfig,mathrsfs,graphicx,rotating}
\usepackage{ifthen}
\usepackage{amsfonts}
\usepackage{ragged2e}
\PassOptionsToPackage{hyphens}{url}
\usepackage[hyphens]{url}
\usepackage{hyperref}
\usepackage{listings}
\usepackage{epstopdf}
\usepackage{color}
\usepackage{float}

\usepackage[normalem]{ulem} 
\usepackage{soul} 
\usepackage{amsmath,amssymb}

\let\originallesssim\lesssim
\let\originalgtrsim\gtrsim

\DeclareRobustCommand{\lesssim}{%
  \mathrel{\mathpalette\lowersim\originallesssim}%
}
\DeclareRobustCommand{\gtrsim}{%
  \mathrel{\mathpalette\lowersim\originalgtrsim}%
}

\makeatletter
\newcommand{\lowersim}[2]{%
  \sbox\z@{$#1<$}%
  \raisebox{-\dimexpr\height-\ht\z@}{$\m@th#1#2$}%
}
\makeatother

\hypersetup{
  colorlinks=true,
  linkcolor=blue,
  citecolor=blue,
  urlcolor=blue
}

\graphicspath{{figs/}}

\newcommand{\beq}{\begin{equation}}
\newcommand{\eeq}{\end{equation}}

\chardef\til=126

\journal{Snowmass 2021}

\begin{document}
\definecolor{mygreen}{rgb}{0,0.6,0} \definecolor{mygray}{rgb}{0.5,0.5,0.5} \definecolor{mymauve}{rgb}{0.58,0,0.82}

\lstset{ %
 backgroundcolor=\color{white},   
 basicstyle=\footnotesize,        
 breakatwhitespace=false,         
 breaklines=true,                 
 captionpos=b,                    
 commentstyle=\color{mygreen},    
 deletekeywords={...},            
 escapeinside={\%*}{*)},          
 extendedchars=true,              
 keepspaces=true,                 
 frame=tb,
 keywordstyle=\color{blue},       
 language=Python,                 
 otherkeywords={*,...},            
 rulecolor=\color{black},         
 showspaces=false,                
 showstringspaces=false,          
 showtabs=false,                  
 stepnumber=2,                    
 stringstyle=\color{mymauve},     
 tabsize=2,                        
 title=\lstname,                   
 numberstyle=\footnotesize,
 basicstyle=\small,
 basewidth={0.5em,0.5em}
}

\begin{frontmatter}

\title{
Jas4pp - a Data-Analysis Framework for Physics and Detector Studies}\tnotetext[mytitlenote]{ANL-HEP-164101, SLAC-PUB-17569}

\author[add1]{S.V.~Chekanov}
\ead{chekanov@anl.gov}

\author[add2]{G.~Gavalian}
\ead{gavalian@jlab.org}

\author[add3]{N.~A.~Graf}
\ead{Norman.Graf@slac.stanford.edu}

\address[add1]{
HEP Division, Argonne National Laboratory,
9700 S.~Cass Avenue,
Argonne, IL 60439, USA.
}

\address[add2]{
Jefferson Laboratory,
12000 Jefferson Ave., Newport News, VA 23602
}

\address[add3]{
SLAC Linear Accelerator Laboratory,
2575 Sand Hill Road,
Menlo Park, CA, 94025
}

\begin{abstract}
This paper describes the Jas4pp framework  for exploring physics cases and for detector-performance studies of future particle collision experiments.
Jas4pp is a multi-platform Java program for 
numeric calculations, scientific visualization in 2D and 3D, storing data in various file formats and  displaying collision events and detector geometries. It also includes complex data-analysis algorithms for  function minimisation, regression analysis, event reconstruction (such as jet reconstruction), limit settings and other libraries widely used in particle physics.
The framework can be used with several
scripting languages, such as Python/Jython, Groovy and JShell.
Several benchmark tests discussed in the paper illustrate significant
improvements in the performance of the Groovy and JShell scripting languages compared to the standard Python implementation in C. The improvements for numeric computations in Java 
are attributed to recent enhancements in the Java Virtual Machine. 
\end{abstract}


\end{frontmatter}

\newpage
\section*{Program summary}

\hspace{-0.6cm}{\it Program title:} Jas4pp \\
{\it CPC Library link to program files:} \\
{\it Developer's repository link:}  \url{https://atlaswww.hep.anl.gov/asc/jas4pp/} \\
{\it Code Ocean capsule:} \\
{\it Licensing  provisions:}\ GNU  General  Public  License  3  (GPL)  \\
{\it Programming  language:} Java, Jython, Groovy \\
{\it Nature of problem:} Develop a platform-independent data-analysis framework
for high-energy and nuclear physics (HEP and NP) with a support of fast dynamically-typed scripting
languages,
comprehensive data-visualisation and I/O libraries. \\
{\it Solution method:}
The solution adopted here is to use Java and the scripting languages integrated with Java VM.\\
{\it Additional comments:}  All 3rd party Java libraries included with this program
are licensed by GPLv3,
GNU Lesser General Public License (LGPL) or by other
licenses compatible with the GPLv3 license, and
adhere to Mendeley Data approved open-source software licenses.
These licenses files  are includes with the program.

\input{intro.tex}

\input{structure.tex}

\input{benchmarks.tex}

\input{usage.tex}

\input{iolibs.tex}

\section{Documentation} 

The Jas4pp package includes example codes implemented in Java, Groovy and Jython. Examples dealing with 2D/3D data visualization can be found from the ``Welcome'' screen of the Jas4pp editor.  More complex examples are located in the directory "examples" of the Jas4pp installation directory.

The API documentation of the main Java classes included with Jas4pp can be accessed from the web site \cite{jas4pp} of this project.

Jas4pp libraries can be used in combination with any advanced  integrated development environment (IDE), such as Eclipse, NetBeans,  IntelliJ IDEA and other IDE. Their setups should point to the directory ``lib'' of the Jas4pp installation. Java reflection used by such IDEs provides the ability to inspect classes, interfaces, enum etc. Such IDEs can also use the HTML documentation of Java classes to provide more information on the implementation of such classes. Note that  introspection of Java classes is also possible
using the Jython shell as described in Sect.~\ref{sec:features}.

It should be pointed out that Jython and Groovy can also be used to list methods of any given object initiated inside a user code.
To print the values stored inside an object, one can use the method ``toString()''.  In Jython,  the method ``dir(obj)'' is used to learn about all the methods of an 
instantiated class ``obj'', following the standard Python approach. Here is the example:

\vspace{0.5cm}

\begin{lstlisting}[language=Python]
from jhplot import H1D
h=H1D("Histogram",10,0,1)
print h.toString() # print what is inside of this histogram
print dir(h)       # print all methods of the Java class jhplot.H1D
\end{lstlisting}
The names of the printed methods are usually self-explanatory. 

In Groovy (or Java, JShell), the method ``toString()'' can also be used, but it is not implemented in a consistent way across all Java classes.  Instead, one can use the method ``inspect()'' to obtain the information about the name of the class and the values of its fields. 

Here is a Groovy example that shows several alternative methods to inspect Java objects:

\vspace{0.5cm}

\begin{lstlisting}[language=Java]
import jhplot.H1D
h=new H1D("Histogram",10,0,1)
println h.inspect()
// access  more information on its values
println h.properties
// Java style to show methods of the class jhplot.H1D
println h.getClass().getMethods() 
\end{lstlisting}

Jas4pp is licensed under the GNU General Public License (GPLv3).
The 3rd party Java libraries included with this program are licensed by GPLv3, 
GNU Lesser General Public License (LGPL) or by other
licenses compatible with the GPLv3 license.

\section{Summary}

This paper describes the Jas4pp framework  \cite{jas4pp} 
designed  for physics and detector studies
of current and future particle-collision experiments.  
Although this framework has been used in several studies 
dealing with physics and detector performance studies, there are no publications 
describing this program. This paper is the first overview of the main features of Jas4pp.

Several examples given in this paper illustrate 
the simplicity of this program for data analysis when using the Jython and Groovy scripting languages.  The Groovy language included with this framework shows a significant improvement in the performance for numeric computations compared to the standard CPython. JShell (from JDK) is another 
fast and friendly environment that can call the Java libraries included with Jas4pp. 
The observed performance improvements in dynamically-typed languages implemented in Java  are 
due to the recent enhancements in the modern JVMs, leading to a significant  increase in 
the speed of evaluations of mathematical functions. This progress will contribute to a wider
usage of Java for numeric calculations. 
In the context of particle physics, this should improve processing speed  for
detector reconstruction software and Monte Carlo simulations that heavily use trigonometric functions.

The Jas4pp program can easily be extended by adding JAR files to
the  ''lib/user'' directory. Such  libraries will be immediately available for user analysis programs  implemented in Java, JShell, Jython/Python and Groovy, without additional modifications for specific operating systems.

Due to its low maintenance, easy installation, convenient programming using scripting languages and high performance for numeric calculations, the Jas4pp program can be a promising computing environment for physics analysis and detector studies of future experiments in particle and nuclear physics.

\section*{Acknowledgments}
We thank Marco Lucchini for help with debugging the Jas4pp program.
We gratefully acknowledge the computing resources provided on a
high-performance computing cluster operated by the
Laboratory Computing Resource Center at Argonne National Laboratory.
The submitted manuscript has been created by UChicago Argonne, LLC,
Operator of Argonne National Laboratory (“Argonne”). Argonne, a U.S.
Department of Energy Office of Science laboratory, is operated under
Contract No. DE-AC02-06CH11357. The U.S. Government retains for itself,
and others acting on its behalf, a paid-up nonexclusive, irrevocable
worldwide license in said article to reproduce, prepare derivative works,
distribute copies to the public, and perform publicly and display
publicly, by or on behalf of the Government.
The Department of Energy will provide public access to these results of
federally sponsored research in accordance with the
DOE Public Access Plan.
\url{http://energy.gov/downloads/doe-public-access-plan}. Argonne
National Laboratory’s work was
funded by the U.S. Department of Energy, Office of High Energy Physics
under contract DE-AC02-06CH11357.

\bibliographystyle{elsarticle-num}
\bibliography{references}

\clearpage
\appendix
\input{examples.tex}

\end{document}

%% file: intro.tex
\section{Introduction}

Software frameworks are central
to physics analysis and detector-performance studies in particle collision experiments. 
In the past decades such analysis frameworks have included PAW (``Physics Analysis Work station'') \cite{BOCK1987181} implemented in C/Fortran,  ROOT \cite{root} written in C++, and Jas3 (``Java Analysis Studio'') \cite{java_toni} implemented in Java. The latter program 
was developed at the SLAC National Accelerator Laboratory. It was
used for the SiD detector concept \cite{Behnke:2013lya} of  the International Linear Collider (ILC) project \cite{Behnke:2013xla}, and then it was extended to a more versatile package with downloadable  plugins for various projects beyond high-energy physics (HEP) experiments. 
For example,  such Java libraries are used in reconstruction, calibration, monitoring and physics analysis of the CLAS12 and HPS experiments \cite{clas12,BALTZELL201769}  in the experimental Hall B
at Thomas Jefferson National Accelerator Facility (Jefferson Lab).

With an increased interest in other future HEP projects, such as CLIC \cite{Linssen:1425915}, the high-energy LHC (HE-LHC),
and $pp$ colliders of the European initiative, FCC-hh~\cite{Benedikt:2206376} and the Chinese initiative (CEPC \cite{CEPCStudyGroup:2018ghi} and  SppC~\cite{Tang:2015qga} experiments), it becomes apparent that sustainable software packages with easy deployment by end-users are important. 
Jas3 was one of the most promising packages to satisfy the above-mentioned sustainability requirement  since it was written in  Java. This programming language has been exceptionally successful in business and enterprise computing since the compilation of Java source code into bytecode  makes it ideal for distributed applications.
As any Java application, Jas3 requires low maintenance and does not have  platform-specific installation issues.

In 2016, a program called Jas4pp (``Jas for particle physics'') based on Jas3 was created at the Argonne National Laboratory (ANL) to accomplish physics and detector performance studies using a SiD-derived reconstruction software for the HepSim  \cite{Chekanov:2014fga} repository. As a Java application,
Jas4pp runs on any platform with Java installed, including Linux, MacOS and Windows OS.  
The time of deployment of Jas4pp is compatible with the time needed to download this program to a local computer since there are no  platform-specific installation requirements. 

This paper gives an introduction to the Jas4pp  program. It will discuss multiple use cases, basic examples that illustrate its use and where to find the needed documentation. 
Section~\ref{benchmarks} will discuss several benchmarks to illustrate the performance of Java and dynamically-typed programming languages included with Jas4pp.
The Appendix  of this paper illustrates several advanced examples with data analysis in particle physics.

%% file: structure.tex
\section{Main features of Jas4pp}
\label{sec:features}

In terms of the software libraries for data analysis in particle physics, 
Jas4pp goes much beyond the PAW and ROOT programs. Jas4pp  contains
a full stack of physics and detector-related libraries integrated with  several programming languages. 

The core part of Jas4pp is the package called Jas3 (Java Analysis Studio) \cite{jas2} developed at SLAC. It is a flexible Java platform for data analysis that can be configured via plugins  for different experiments. Jas3 is based on JAIDA \cite{Donszelmann:2008zz},  
a Java implementation of the Abstract Interfaces for Data Analysis (AIDA). 

Being backward compatible with the original Jas3, Jas4pp focuses on data analysis in collider particle physics. 
One distinct feature of Jas4pp is that it includes HEP libraries by default instead of requiring additional plugins to be downloaded. This makes Jas4pp a self-contained program for analysis and event visualization for particle-collision experiments.

The Jas4pp program consists of:  (a)  2D/3D scientific visualization libraries, similar to ROOT and PAW,
(b) data containers and numerical libraries (including  non-linear regressions using an interactive GUI), (c) physics libraries with Lorentz vectors, limit setting, event-shape studies and jet reconstruction. In particular, the anti-$k_T$ algorithm \cite{Cacciari:2008gp}  for analysis of $pp$ events is included, following the algorithmic solutions implemented in  the C++ FastJet package~\cite{Cacciari:2011ma}.
(c) A full-featured event display and a data-container browser which 
can be used to visualize detectors and study collision events after the Geant4 simulation \cite{Allison2016186}. 
Finally, Jas4pp can be used to analyze truth-level events from the HepSim repository \cite{Chekanov:2014fga} which contains more than 100 scripts executed directly in Jas4pp.

Jas4pp supports three main programming languages for analysis code: Java, Python and Groovy. The latter two are implemented in the Java libraries that come with  the Jas4pp package itself. 
Python and Groovy are  dynamically-typed scripting languages,
to be executed on the Java platform. With the increased use of the Python language in HEP, Jas4pp adopted Jython (version 2.7.2) as the main programming language for user analysis.  This version 
is compatible with  CPython 2.7.2 (implemented in C).
It is supported in the GUI mode (using the built-in editor 
with the console for outputs), as well as using
the batch (console)  mode. In addition, Apache Groovy \cite{groovy} (version 3.0.6) is supported. Similar to
Jython, Groovy is an optionally typed dynamically-typed language. It allows smooth integration with
Java and any third-party library.
The main advantage of Groovy is that the execution speed
is significantly faster than that for the equivalent Jython or CPython codes.
Benchmarks indicate that execution of Groovy code that implements long loops is more than a factor ten faster than the interpretation of the equivalent code implemented in Jython / CPython. Such benchmarks will be discussed in Sect.~\ref{benchmarks}.

Compared to the original Jas3, the Jas4pp program includes additional Java libraries from the DataMelt platform \cite{chekanovbook},   GROOT project \cite{groot} developed at JLab, LCIO \cite{Gaede2003LCIOA} and improved LCSIM \cite{6551260} libraries \footnote{The LCIO and LCSIM libraries were included in Jas3 as downloadable plugins.}, mathematical libraries from  the Apache foundation and a software library \cite{Chekanov:2015cca} for dealing with  Monte Carlo
event files from the HepSim repository \cite{Chekanov:2014fga}. 
It also includes several other additional libraries for versatile visualisation of data.

Jas4pp  supports a number of I/O file formats discussed in Sect.~\ref{secIO}.
In addition to the traditional LCIO file format  \cite{Gaede2003LCIOA}, 
Jas4pp supports several other new data formats used in different projects.

The LCSIM library included in Jas4pp was significantly improved to increase the processing speed of tracking hits when complex events are visualized inside Jas4pp event display. This was achieved by replacing the standard Java trigonometric functions from the "Math" package with their fast implementation using the FastMath Java package \cite{fastmath}. The result of this replacement is a factor five faster rendering of  objects in "busy" events with a large number of reconstructed tracks when using previous versions of Java (version 8 and below).

\begin{figure}[ht]
  \centering
    \includegraphics[width=0.9\textwidth]{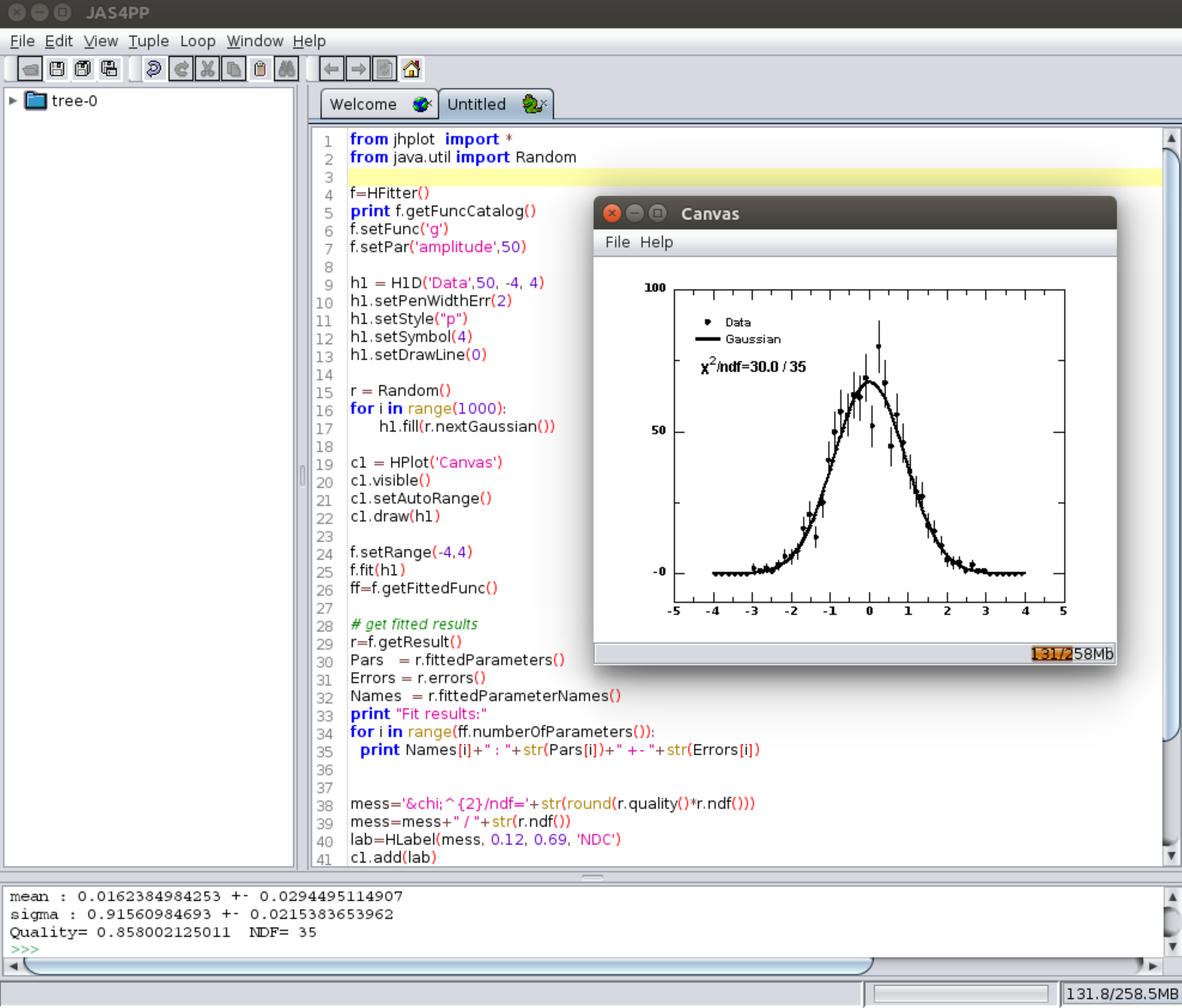}
    \caption{An illustration of the Jas4pp editor with the result of the code execution. The program is implemented in Jython.}
  \label{fig:jas4pp_editor}
\end{figure}

Jas4pp has a full-featured editor with syntax highlighting (a feature that is missing in the original Jas3 package). Figure~\ref{fig:jas4pp_editor} illustrates the editor with an analysis code written in the Python language, and the result of the execution of the code illustrating a fit of the Gaussian-distributed data.

\section{Running Jas4pp}
\label{sec:run}

Jas4pp runs on any operating system that supports the Java Virtual Machine (JVM) and has been explicitly tested  on Linux, MacOS and Windows 10.
Typically, 4 GB of RAM is sufficient.
Jas4pp (version 1.5 released in October 2020) requires Java SE Development Kit 8 (JDK 8) and above.
Jas4pp was also tested with OpenJDK 9 and 14. 

After downloading the Jas4pp compressed tar file from the official web site \cite{jas4pp},
the program should be extracted to the directory ``jas4pp''.
To start the GUI editor shown in Fig.~\ref{fig:jas4pp_editor}, the following command should be executed in the Linux console:

\vspace{0.5cm}
\begin{lstlisting}[language=Python]
bash> source ./setup.sh # setup the environment
bash> jaspp             # start the Jas4pp editor 
\end{lstlisting}
or using the equivalent ``jaspp.bat'' script for the Windows OS. The last command can take an arguments (such as a name of the file to be opened).

To process a Jython or Groovy script in batch mode, the following command should be invoked  using "example.py" (Jython) or "example.groovy" (Groovy) file with an analysis code:

\vspace{0.5cm}
\begin{lstlisting}[language=Python]
bash> fpad example.py      # process a Jython script
bash> fpad example.groovy  # process a Groovy script
\end{lstlisting}
Note that Groovy can use the alternative file extensions "gvy" and "gy".

The Jas4pp editor can also be used to edit programs that use 
the Java or JShell syntax\footnote{JShell is an interactive tool for  prototyping Java code. It comes together with the JDK installation.}.

Another useful command is ``fpad\_edit''.  It is a light-weight editor for Jython code. The editor includes an interactive Jython console with a built-in help system where the keyboard shortcut [CTRL]+[SPACE] after the dot displays all methods for any instantiated object/class, i.e:

\vspace{0.5cm}
\begin{lstlisting}[language=Python]
>>> a="This is a Python string. It can be a Java class too"
>>> a. # press [CTRL]+[SPACE] to display all methods of the object "a"
\end{lstlisting}

A convenient way to compile and run  Java programs using the Linux console is to add all JAR (Java archive) 
files from the directory ``lib'' of 
the Jas4pp installation directory to the Java CLASSPATH. 
The same method can be used for JShell.
Examples of such approaches are available in dedicated books on Java.

Another important directory is called "examples". It includes code snippets that illustrate various aspects of Jas4pp computing. The code examples are implemented in the Jython and Groovy scripting languages. These examples can straightforwardly be converted into  the  Java and JShell programming styles of analysis programs.

%% file: benchmarks.tex
\section{Benchmarks}
\label{benchmarks}

Before the introduction of  the just-in-time (JIT) compiler, 
Java was only interpreted and not compiled, and thus 
programs implemented in Java were slow compared to C/C++. Since the introduction of JIT, the performance of the Java Virtual Machine (JVM) has improved over the years. Even though there are several  aspects of Java that are very appealing for the purposes of physics analysis software there is still a reservation in the physics community whether Java can be used for writing analysis code.  Java produces platform-independent libraries that can be distributed easier than large scale C++ frameworks, which require compilation for each machine the analysis code is deployed to. The reflection mechanism in Java allows writing plugins based on run-time configurable workflows.

When Java is discussed in the context of data analysis, the main reservation is the performance of the JVM compared to other widely used languages in nuclear and high energy physics 
such as C, C++ and FORTRAN.
Slow evaluation of mathematical functions and large memory footprint were the most 
significant drawbacks for numeric computations of the previous versions of JVM.  
Trigonometric functions are especially important for track reconstruction software 
in particle physics and for numeric calculations in general.
Recent enhancements in JVM enable faster evaluations of mathematical expressions in large
loops. Some aspects of the JVM performance improvement using an example based on the  
``java.lang.Math'' package  will be discussed later in this section.

\subsection{Java vs other languages}
\label{sub:ben1}

This section will discuss a few benchmark tests implemented in  different programming 
languages to measured computation time. 

In the first test, we compute the value of $\pi$ using a  Monte Carlo method.
The code that implements this calculation is illustrated using 
the Groovy scripting language:

\begin{lstlisting}[caption={Groovy code for the calculation of $\pi$.},label={groovy},language=Java]
import java.util.Random
int nTh = 0; int nSu = 0
double x, y
then = System.nanoTime()
r = new Random()
for (int i = 0; i < (int)1e8; i++) {
   x = r.nextFloat(); y = r.nextFloat()
   nTh++
   if ( x*x + y*y <= 1 )  nSu++ }
itime = ((System.nanoTime() - then)/1e9)
println "Time: "+itime+" sec, Pi="+4*nSu/(double)nTh
\end{lstlisting}
To run this code using Jas4pp, save these lines in a file with the extension "groovy", "gvy" or "gy", and process it
as described in Sect.~\ref{sec:run}.

\begin{figure}[ht]
  \centering
    \includegraphics[width=0.8\textwidth]{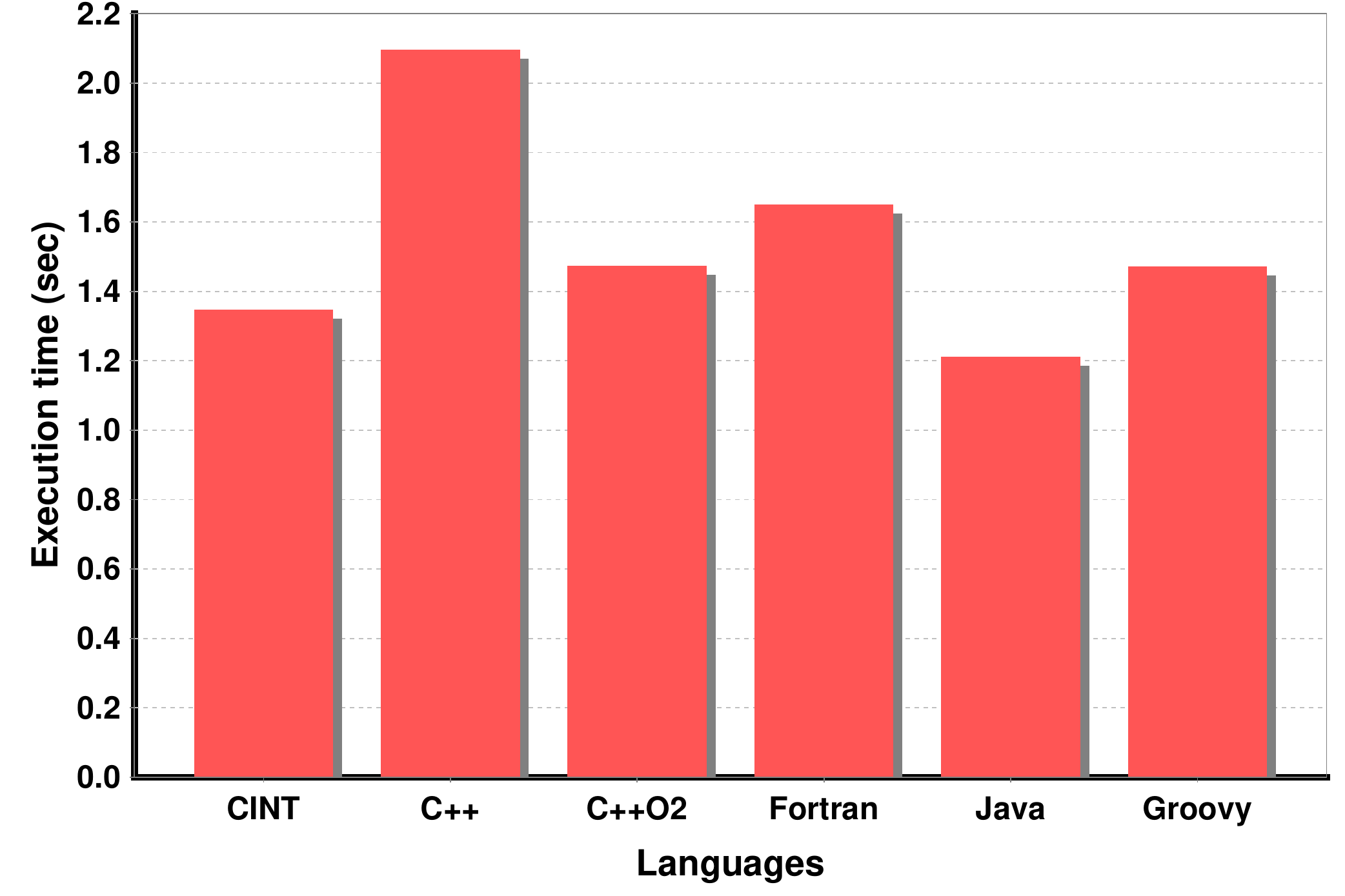}
    \caption{Calculations of $\pi$ using Monte-Carlo method, using random number generator. The plot was obtained using the Jython example in \ref{app0}.}
  \label{fig:benchmark_pi}
\end{figure}

The results of this test are shown in Figure~\ref{fig:benchmark_pi}. 
Java (running on OpenJDK 11 x64) performed better than C++ and FORTRAN for this test.  
These benchmarks are performed on an AMD EPYC 7502 (32-core) machine, using single-threaded code. The time measured was for $10^8$ iterations. CPython (Python implemented in C) was also used in the benchmarks, but it is  omitted from the graph because of the large execution time (75 seconds). 
This image was programmed using Jython and the Jas4pp program as shown in  \ref{app0}.  

\begin{figure}[ht]
  \centering
    \includegraphics[width=0.9\textwidth]{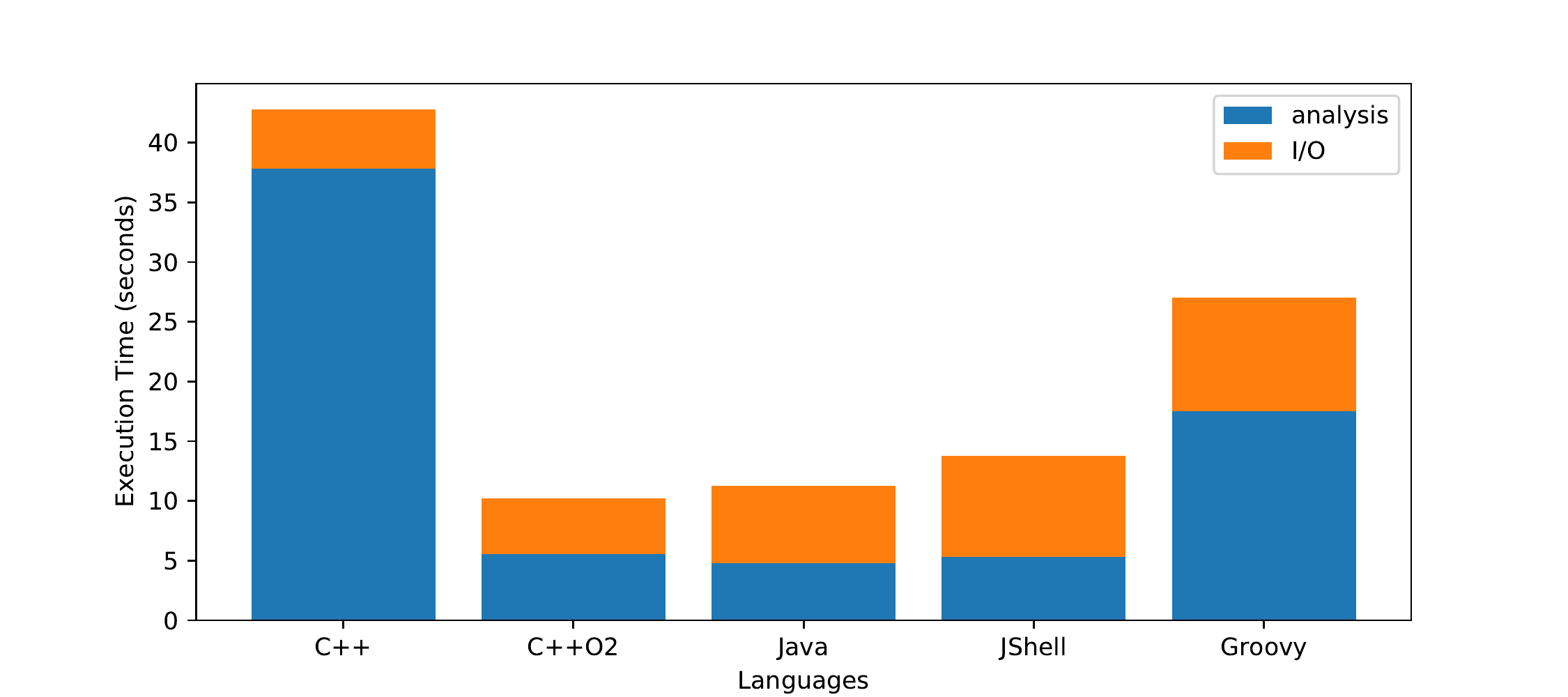}
    \caption{Benchmark of physics analysis codes similar to typical analysis done 
    in nuclear physics.}
  \label{fig:benchmark_ana}
\end{figure}

In the second test, we ran a simple analysis code, on real experimental data, where particle and corresponding detector information is read from a data stream and photons are reconstructed and counted based on their response in an electromagnetic calorimeter. The same code was written in C++ and Java, and the Java code was run through JShell and Groovy to measure the performances of Java's scripting environments. The results of this test are shown in Figure~\ref{fig:benchmark_ana}.

The test shows that the performance of Java is very similar to the C++ code performance. What is  interesting in these tests is that JShell, which is the Java interactive shell provided with JDKs starting from version 9, performs very well for an interpreted scripting language. It is worth mentioning that Java shows  the best performance when there is complex Object Oriented Code, where the JIT compiler is most effective in identifying hot spots and in-lining function calls during run-time to achieve the best performance.

Designing and testing large-scale applications in many languages is difficult, but even these simple tests indicate that over the years Java performance has significantly improved and can be considered as a good choice for writing data analysis software for nuclear and high energy physics.

\subsection{Jas4pp-supported scripting languages}
\label{sub:ben2}

Jas4pp supports the Java, Python/Jython and Groovy programming languages. 
Comparisons of the execution time
for Java and Groovy were shown in the previous section. In this section we will
discuss the performance of Jython (Python)  and Groovy due to the  importance
of Python in the modern data-analysis environment.

The code used to calculate the value of $\pi$  using a Monte Carlo technique can be written in Python as follows:

\begin{lstlisting}[caption={Python code for the calculation of $\pi$.},label={python2},language=Python]
import time,random
nTh,nSu = 0,0
then = time.time()
for i in xrange(int(1e8)):
  x,y = random.random(),random.random()
  nTh +=1
  if ( x*x + y*y <= 1 ):  nSu+=1
itime = time.time() - then
print ("Time ",itime," sec, Pi=",4*nSu/float(nTh))
\end{lstlisting}
To run this code using Jas4pp, save these lines in a file with the extension "py" and process it
as described in Sect.~\ref{sec:run}. The processing time for the above calculation in Jas4pp is 36 seconds using Jython 2.7.2 employing
the same CPU and JDK as for the previous tests.  40 seconds was obtained for CPython 2.7.2, and 46 seconds for CPython 3.4.3  (after replacing the function {\tt xrange()} with {\tt range()}).
The  PyPy interpreter for the Python language required about 5 seconds for the same code.

The same algorithm implemented using Groovy was discussed before (see Listing~\ref{groovy}). The execution speed for Groovy was 4 seconds. Thus the gain in the speed of Groovy 
can be as large as a factor 10 compared to CPython.
Removing the explicit declarations (int, float) at the beginning of the Groovy code leads to an execution time of about 20 seconds.

The benchmarks were repeated using OpenJDK 13 (x64) and an Intel(R) Core(TM) i5-4690K CPU @ 3.50GHz.
Although this combination of such CPU and JDK showed overall better performance for the benchmark codes, the  conclusion about the relative  difference in  execution time for different scripting languages was similar to the tests based on the AMD EPYC CPU.

In conclusion, Jython execution of the benchmark code  is similar to CPython2 (implemented in C).  
The performance of CPython3 interpreter is surprisingly worse than CPython2, and is significantly worse than for Jython. The Groovy scripting language  supported by Jas4pp showered a significant improvement  in execution time compared to the equivalent code implemented in the Python language (CPython or Jython). The improvement factor depends on how variables are defined (explicit vs implicit declarations).
The performance of Groovy is also better than for the PyPy interpreter.

\subsection{Evaluation of trigonometric functions}
\label{sub:ben3}

It should be noted that our conclusion about the speed of Python code execution in the previous section 
is obtained using the specific algorithm, i.e. the evaluation of the value of $\pi$ using a Monte Carlo technique. It is possible that  a different result may be obtained for other benchmark algorithms or different implementations. However, our example is quite common for a typical numeric calculation involving random numbers and non-linear transformations. 

Fast implementation of trigonometric functions is one of the areas where the 
Java-implemented Python interpreter may still be behind C/C++. There is no immediate answer to the question on  
code execution speed for every possible situation since such benchmarks  depend on specific algorithms.
For example, Listing~\ref{python3}
shows the situation when processing speed of Jython is about a factor 2 slower than for CPython.

\vspace{0.5cm}
\begin{lstlisting}[caption={Python code with  trigonometric functions},label={python3},language=Python]
import math,time
then = time.time()
x=0
for i in xrange(int(1e8)):
   x=x+math.sin(i)/math.cos(i)
itime = time.time() - then
print("Time:",itime," (sec) result=",x)
\end{lstlisting}

The above code was processed using Jas4pp with OpenJDK 13 (x64). The same algorithm re-written in the Groovy language shown in Listing~\ref{groovy3} is a 10\% (30\%) faster than for CPython2 (CPython3).

\vspace{0.5cm}
\begin{lstlisting}[caption={Same code as in Listing~\ref{python3} but implemented in Groovy},label={groovy3},language=Java]
import java.lang.Math
long then = System.nanoTime()
double x=0
for (int i = 0; i < 1e8; i++)
    x=x+Math.sin(i)/Math.cos(i)
itime = ((System.nanoTime() - then)/1e9)
println "Time: " + itime+" (sec) result="+x
\end{lstlisting}

The same  algorithm re-implemented in Java and processed using OpenJDK 13 (x64) further increases the execution speed by a factor 2 compared to the Groovy dynamic language. 

Similar benchmarks of the Java code have been carried out by repeating the calculation using Java SE 8 (JDK 1.8 from Oracle) released in March 2014, and using Linux Ubuntu 20.04 LTS (x64).
The computation on the JDK 1.8 is about a factor 8 slower compared to OpenJDK 13. The difference in the speed is reduced from 8 to 3 for the AMD EPYC CPU, indicating that the improvement in speed depends on the CPU architecture. Similar improvements in the speed compared to the JDK 1.8 were observed for other trigonometric functions of the ``java.lang.Math" package.

Thus the  benchmark tests illustrate significant progress in effectiveness of 
modern JVM for numeric computations.
It was found that  OpenJDK 11 and above  has  a performance similar to  the ``FastMath'' package from the Apache Common Math project \cite{fastmath} or Jafama \cite{jafama}. They  heavily rely on optimizing compilers to native code, and use of large tables for mathematical functions.  Therefore, even applications compiled to the bytecode  using older Java versions should show improvements for large-scale numeric computation on modern JVM. This progress in the performance of JVM is expected to contribute to speed improvements for Monte Carlo simulations and event reconstruction software used in nuclear and particle physics that heavily utilize loops with trigonometric functions.

The algorithm shown in Listing~\ref{groovy3}
but implemented in C++ was about 30\% faster than for the Java code processed with  OpenJDK 11 (13).
The C++ code was compiled using GCC 9.3 on Ubuntu 20.04.
Although the compiled C++ code shows a better overall performance, the  difference
between C++ and OpenJDK 11/13 is significantly smaller than for C++ vs JDK 1.8.
For end-user analysis, several tens of percent slower processing speed  of Java  compared to C++ is a small price to pay when it comes to the Java  advanced features, such as user friendliness, platform independence, automatic memory management, built-in multithreading support and the reflection technology.

The question of code profiling using different implementations is a complex problem, 
and we do not plan to explore all possible scenarios for this article.
The main conclusion we want to draw in this section is that the processing speed of 
the code that implements large loops with numeric calculations is substantially better for 
Groovy than for CPython (or Jython). The speed improvement for Groovy (which is closer integrated with Java than Jython) is attributed to the recent performance enhancements in JVM. 

We should remind that the execution speed of a CPython code can significantly be increased using NumPy \cite{harris2020array} or similar external libraries implemented in C/C++ and compiled to object files. 
This question  is outside the  scope of this section dealing with implementations of analysis code in dynamically typed languages. However, here we should mention that  applications used in particle and nuclear  physics typically deal with manipulations of complex custom-designed objects. Therefore, the usage of  general third-party libraries called by CPython, such as NumPy, is limited. 
In the situation when the bulk of the calculations are implemented
in external libraries, the  speed of algorithms implemented in C/C++ libraries should be compared with the speed of execution of Java bytecode libraries (see the discussion in Sect.\ref{sub:ben1}).  The latter can be called from  Jython, Groovy or JShell in the same way as CPython calls 3rt-party libraries implemented in C/C++.

%% file: usage.tex
\section{Usage of Jas4pp}

As mentioned in the introduction,
Jas3 was extensively used for the SiD detector concept \cite{Behnke:2013lya} of the ILC project.
Jas4pp still maintains the required libraries for analysis of collision events created by the ILC community. In particular, Jas4pp natively reads the "miniDST" events in the LCIO file format used for $e^+e^-$ studies. Such examples are  available from the Jas4pp web page.

Jas4pp has also been used in several other  detector studies focused on future experiments.
For example, it was used for  designing  a silicon tracker for the future Circular Electron Positron Collider (CEPC) experiment \cite{CEPCStudyGroup:2018ghi}, as an alternative option
to the time projection chamber (the so-called ``TPC'') tracker.  Calorimeter studies for the FCC-hh~\cite{Chekanov:2016ppq,Yeh:2018ujb, Yeh:2019xbj,Chekanov:2020xco} future experiment were largely conducted using this program.
The Java libraries included in Jas4pp  were  used for studies of  crystal calorimeters for future lepton colliders \cite{lucchini2020new}.
Jas4pp  was the main framework for the initial development of the TOPSiDE detector of the 
Electron-Ion Collider (EIC)  collider~\cite{Repond:2018kap}. 
It was also  used for HL-LHC and HE-LHC studies \cite{Chekanov:2018nuh} based on HepSim Monte Carlo event samples. 

The Java libraries included with Jas4pp libraries were also used for several ongoing experiments. For example, the plotting libraries included with Jas4pp were used for detector studies of the  hadronic calorimeter and the event display of the ATLAS experiment \cite{Collaboration_2008} at the LHC.  The libraries (such as the ones that come with the GROOT package) are currently used at Jefferson Laboratory for monitoring, reconstruction, calibration and physics analysis of the CLAS12 and HPS experiments \cite{clas12,BALTZELL201769}. 

Applications that use graphics and numeric computations implemented in C/C++ are especially challenging for maintenance over many years due to changes in the C/C++ compiled libraries included 
with the Linux operating system used in particle physics.
This is much less of an issue for Java. As a Java package,  Jas4pp is expected to require low maintenance over the years. It was verified that version 0.4 of the Jas3 program developed in 2003 can run on modern computers with OpenJDK 14 (released in March 2020) without problems. Thus, Java-based end-user environments can be the optimal solution for experiments to be built  in 20-30 years from now due to strong backward compatibility of Java and low requirements for maintenance of this application.

The following section will discuss different uses of Jas4pp for physics analysis and detector visualization.

\subsection{Validation of Monte Carlo events}

Jas4pp can be used for  validation  of Monte Carlo generators as discussed in \cite{Chekanov:2015cca}.
This functionality includes running analysis scripts and checking histograms with entries
of different kinematic quantities. In particular, it allows  jet algorithms to be run on input Monte Carlo data from the HepSim repository \cite{Chekanov:2014fga}. 
The coding can be implemented in either Java, Groovy or Jython. Since this topic was extensively discussed in Ref.\cite{Chekanov:2015cca}, we will skip any further discussion.

\subsection{Browsing event containers}

Jas4pp, as its predecessor, Jas3, can be used to browse event containers stored in the LCIO files.
This functionality is almost unchanged in Jas4pp program.

\subsection{Interactive fit of data}

Jas4pp, as its predecessor, Jas3, can be used for interactive
fitting of data. This feature is quite unique: a user can adjust initial parameters of fit functions by dragging the mouse pointer, before a minimization operation is applied. 
This significantly simplifies fitting complex functions
since the initial parameters can be set visually following the shape of the data. This functionality is almost unchanged in the Jas4pp program. In addition to the standard method of fitting, Jas4pp allows calling the interactive fit program directly from user analysis codes.
Some examples of such fits are given in the directory ''examples''.

\subsection{Visualisation of detector geometry}

Jas4pp preserves the standard Jas3 method to visualise detector
geometry in 2D or 3D. It can open HepRep geometry files \cite{Perl2000HepRepAG} using the menu [File]-[Open data source]-[HepRep]. A collection of HepRep geometry files
for different collider experiments can be found in the HepSim repository.

\subsection{Event display}

\begin{figure}[ht]
  \centering
    \includegraphics[width=0.9\textwidth]{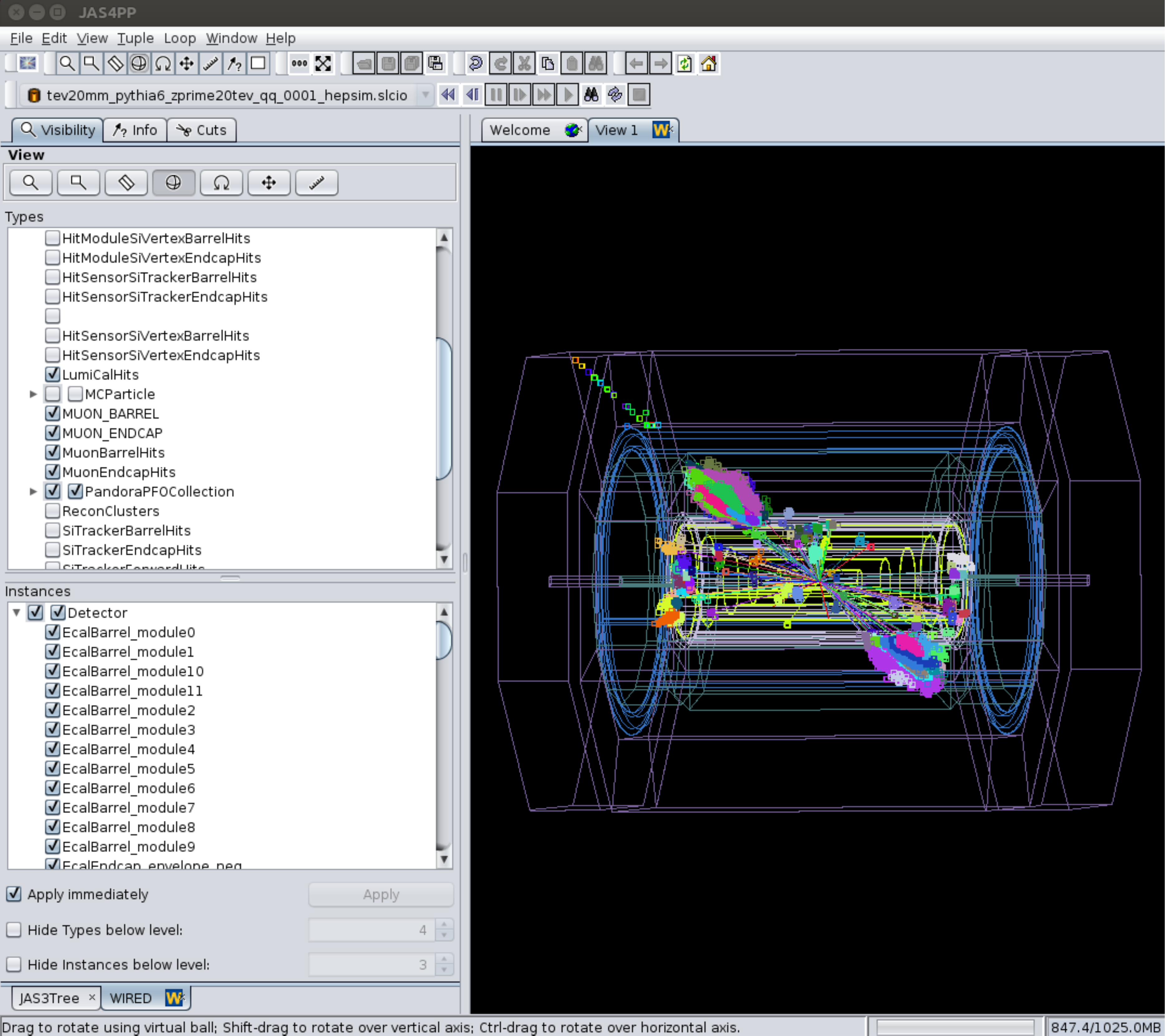}
    \caption{An illustration of the Jas4pp event display of a collision event with the process $Z'\to q\bar{q} \to 2\> \mathrm{jets}$, assuming the mass of 20 TeV for $Z'$. The event display corresponds to the SiFCC detector \cite{Chekanov:2016ppq}.}
  \label{fig:jas4pp_display}
\end{figure}

Event display is another feature inherited from Jas3. The original implementation 
is based on the Wired4 event display \cite{BALLAMINUT2001266,Donszelmann:865605}. It is used for event visualization of data stored inside LCIO files. The event display relies on the LCSIM java reconstruction software, which was improved in terms of speed by replacing the trigonometric functions used for tracking with a fast implementation from the Apache Common Math library. As the result of this, the speed for event rendering has significantly improved. 

Jas4pp creates detector rendering graphics  after opening a LCIO file with truth-level and detector-level information,
and downloading detector geometry files based on the name of the detector simulation used to create the LCIO file.
Unlike Jas3, Jas4pp downloads detector geometries from the HepSim repository.

Figure~\ref{fig:jas4pp_display} shows a complex collision event of two high-energy muons. The centre-of-mass energy of the collision is sufficient to create a hypothetical particle $Z'$ with the
mass of 20 TeV decaying to a pair of quarks ($q\bar{q}$), which are then decaying to two hadronic jets. The event display shows the particle-flow objects (created from charged tracks and calorimeter clusters) and an outgoing reconstructed muon crossing the muon detector. Other detector-level objects (tracking hits, clusters, etc.) are deselected for better visibility. The detector geometry corresponds to the SiFCC detector \cite{Chekanov:2016ppq}.

\subsection{Data analysis using  scripting languages}

Jas4pp supports different styles for data analysis, such as (1) JAIDA style using Java factories; (2) DataMelt style that uses short notations for histograms and arrays; (3) GROOT style that resembles the pyROOT syntax.
In order to illustrate these different styles of programming, we will give three code examples
that show how to fill a 1D histogram with random numbers using a Gaussian distribution:

{\it Using the JAIDA style \cite{Donszelmann:2008zz}:}

\begin{lstlisting}[language=Python]
from hep.aida import *
from java.util import *
fac = IAnalysisFactory.create()
hf = fac.createHistogramFactory(None)
h1 = hf.createHistogram1D("Example",10,-2,2)
r = Random()
for i in xrange(100):
    h1.fill(r.nextGaussian())
c1 = fac.createPlotterFactory().create("plot")
c1.show()
c1.createRegions(1,1)
c1.region(0).plot(h1)
\end{lstlisting}

{\it Using the DataMelt style \cite{chekanovbook}:}

\begin{lstlisting}[language=Python]
from jhplot  import *
from java.util import *
c1 = HPlot()
c1.visible()
c1.setAutoRange()
h1 = H1D("Example",10,-2,2)
rand = Random()
for i in xrange(100):
   h1.fill(rand.nextGaussian())
c1.draw(h1)
\end{lstlisting}

{\it Using the GROOT style \cite{groot}:}

\begin{lstlisting}[language=Python]
from org.jlab.groot.data import *
from org.jlab.groot.ui import *
from java.util import *
c1 = TCanvas('c',500,500)
h1 = H1F('h',100,-5.0,5.0)
rand = Random()
for i in xrange(100):
    h1.fill(rand.nextGaussian())
c1.draw(h1)
\end{lstlisting}
The latter code has a large  similarity with the pyROOT equivalent code
(but the packages are imported from the GROOT Java library). 

Another data-plotting example  used to create a chart shown in  Fig.~\ref{fig:benchmark_pi} 
can be found in \ref{app0}.
More examples with data visualisation that cover histograms, data shown as symbols and error bars, 2D scatter plots and density plots,  can be found in the directory ``examples'' of the Jas4pp installation.

Jas4pp enables analysis code dealing with reconstructed events to be easily written.
All such examples are located in the directory ``examples'' of the 
installation package. \ref{app} gives code snippets showing how to analyse 
 LCIO files containing Monte Carlo events. \ref{app1} shows how to
run over an LCIO file and extract truth-level information from the "MCparticle" table
used to keep particles created by Monte Carlo generators.
This example also shows how to fill a histogram with the $p_z$ momentum component of 
particles and display it in a canvas. 
\ref{app2} shows how to process reconstructed tracks from collision events, and fill
Lorentz-vector particle objects. \ref{app3} shows how to create anti-$K_T$ jets from calorimeter clusters.
The execution speed for these examples is typically slower by a factor of 3-6 than the equivalent codes implemented in Java since Jython is a scripting language.

%% file: iolibs.tex
\section{Supported data formats}
\label{secIO}

Similar to Jas3, Jas4pp fully supports the LCIO \cite{Gaede2003LCIOA} I/O library developed for ILC studies. Some examples of reading LCIO files using Jython code can be found in \ref{app1} (and in the following sections). 

There are many data formats used for storing experimental data by different experiments. Usually the file format changes from the data acquisition stage to final data summary tapes (DST) output for physics analysis. This can represent some challenges when data is translated from one format to another spending empty CPU cycles. A new data format was developed in Hall-B (Jefferson Lab, CLAS12 Detector) to address these issues. The High Performance Output (HiPO) data format \cite{hipo4} was developed to be used in all stages of data processing, starting from raw data going to DSTs. The main advantage of the HiPO data format is a fully indexed layout that allows chunk reading from disks, avoiding many IOPs, and the data structure (record based) is very well suited for parallel data processing. HiPO is using the LZ4 data compression algorithm, which is the fastest at present. It also supports GZIP if needed. 
In recent developments the HiPO library was also complemented with an XRootD \cite{xrootd} driver allowing  HiPO files to be read from XRootD servers. This Java-based XRootD protocol was developed at JLab  since there were no existing Java libraries supporting the XRootD client functionality. To accommodate the diverse needs of the CLAS12 collaboration, the HiPO library was also implemented in C++ and FORTRAN. 

The HiPO and XRootD protocol  libraries are fully available in the Jas4pp distribution via the GROOT library. Some examples of the usage of such libraries can be found in the ``examples'' directory.

Jas4pp also supports the ProMC \cite{Chekanov:2013mma} and ProIO \cite{osti_1513238} file formats based on the Google Protocol buffers library where data can be encoded using Google's platform-neutral, extensible mechanism for serializing structured data. These data formats can include the ``varints'' types that allow serializing integers using one or more bytes. The ProMC and ProIO libraries
are used in the  HepSim repository \cite{Chekanov:2014fga} with Monte Carlo files.
Jas4pp can stream ProMC data over the https protocol. A number of such examples can be found
in the "examples" directory of  Jas4pp installation.

Jas4pp can be used to store  data in a serialized form (in XML or binary formats), leveraging the native Java or Python API. Many Jas4pp objects extend the standard Java "Serializable" interface. This feature enables the straightforward storage of many complex objects using the 
standard Java serialization mechanism.
A simple Jython/Python example of such an approach using the DataMelt "wrappers" class ``HFile''  is given below:
\vspace{0.5cm}

\begin{lstlisting}[language=Python]
from jhplot import H1D
from jhplot.io import HFile
h=H1D("Histogram",10,0,1)
f=HFile("test.ser","w") 
f.write(h)
f.close()
\end{lstlisting}
In this example, a histogram object ``h'' is serialized to a compressed file ``test.ser''. 
Changing ''HFile'' to ''HFileXML'' will allow storing this histogram object in a human-readable XML file. The histogram can be read back using the same classes and the method ``read()''.  The method allows to add keys to access particular objects.

Finally, Jas4pp has  a limited support for reading ROOT files. Currently, only a few basic ROOT structures (histograms and TGraph) can be imported by Jas4pp.

%% file: examples.tex
\section{Examples of JAS4pp analysis codes}
\label{app}
In this section we will consider several examples of reading LCIO files and accessing stored containers.
If you test these examples without using the Jas4pp editor, you should setup Jas4pp on Linux/Mac
with "bash setup.sh" and then run "fpad script.py". If you use
the Jas4pp GUI, copy and paste the code shown below, save it in a file with the extension ".py" 
and run it using the pop up window that appears after clicking the mouse button.

All these examples are written using the Python syntax. 
However, they can be trivially re-written in Groovy, JShell or Java.

\subsection{A simple chart}
\label{app0}

This example implemented in Jython is used to create the bar chart shown in Fig.~\ref{fig:benchmark_pi}. 
It uses the DataMelt Java class (HChart) and the JFreeChart library \cite{jfreechart} included with the Jas4pp program. 
When setting the bar values, we use the category "1" (can be an arbitrary string). This example also illustrates 
 how to export the image to a PDF file. If you need to highlight a separate bar, replace "1" with any string.

\vspace{0.5cm}
\begin{lstlisting}[language=Python]
from jhplot import HChart
c1 = HChart("")
c1.visible()
c1.setNameX("Languages")
c1.setNameY("Execution time (sec)")
c1.setChartBar()
xaxis = ["CINT","C++","C++O2","Fortran","Java","Groovy"]
data =  [1.348, 2.096, 1.474,  1.65,    1.211, 1.472]
for i in range(len(xaxis)):
       c1.valueBar(data[i], xaxis[i], "1")
c1.update()
c1.setLegend(False)
c1.export("pi_results.pdf")
\end{lstlisting}

\newpage
\subsection{Running over truth-level Monte Carlo events}
\label{app1}

As a simple example of Jas4pp usage, we will illustrate how to  read truth-level information
on particles created by Monte Carlo simulations. Here is an example of how to 
extract the "MCParticle" container, loop over events, and fill a histogram with the $z$ component of momenta.  The input file used by this example can be downloaded  from the HepSim repository \cite{Chekanov:2014fga}. This code is written in Jython. As usual in Python, use the method ``type(obj)'' and ``dir(obj)'' to learn about the type of the object ``obj'' and its methods.

\vspace{0.2cm}

\begin{lstlisting}[language=Python]
from hep.lcio.implementation.io import LCFactory
from jhplot import H1D,HPlot  # import graphics
files=["gev250ee_pythia6_zpole_ee.slcio"]
factory = LCFactory.getInstance()
nEvent=0
h1=H1D("Mass",70,50,120)
for f in files:
  print "Open file=",f
  reader = factory.createLCReader()
  reader.open(f)
  while(1):
    evt=reader.readNextEvent()
    if (evt == None): break
    nEvent=nEvent+1
    print " file event: ",evt.getEventNumber(), " run=",evt.getRunNumber()
    col = evt.getCollection("MCParticle")
    nMc=col.getNumberOfElements()
    for i in range(nMc): # loop over all particles 
      par=col.getElementAt(i)
      if(par.getGeneratorStatus() == 1 and par.getCharge() !=0):
          vertex = par.getVertex();
          pdg=par.getPDG()
          momentum = par.getMomentum()
          ee=par.getEnergy()
          mass=par.getMass()
          px,py,pz=momentum[0],momentum[1], momentum[2]
          h1.fill(pz)
    del col,evt
  reader.close() # close the file
  del reader
 
c1=HPlot()
c1.visible()
c1.setAutoRange()
c1.setMarginLeft(100)
c1.setNameX("Pz [GeV]")
c1.setNameY("Events")
c1.draw(h1)
c1.export("mc_truth.pdf")
\end{lstlisting}

\newpage
\subsection{Running over reconstructed tracks}
\label{app2}

Another example deals with an analysis of reconstructed tracks.
The LCIO input files used by this example can be downloaded from the HepSim repository \cite{Chekanov:2014fga}. Unlike the previous example, this code runs over a number of files
in a directory. It accesses ``track'' containers, and prints the 4-momentum
of tracks using the 4-Lorentz vector (based on the LParticle Java class).

\vspace{0.2cm}
\begin{lstlisting}
from hep.lcio.implementation.io import LCFactory
from org.lcsim.event.base import BaseTrackState
from hephysics.particle import LParticle
from math import sqrt
import glob
files=glob.glob("files/*.slcio")
factory = LCFactory.getInstance()
nEvent=0
for f in files:
  print "Open file=",f
  reader = factory.createLCReader()
  reader.open(f)
  while(1):
     evt=reader.readNextEvent()
     if (evt == None): break
     nEvent=nEvent+1
     # print " file event: ",evt.getEventNumber(), " run=",evt.getRunNumber()
     if (nEvent%100==0): print "# Event: ",nEvent
     strVec = evt.getCollectionNames()
     if nEvent == 1:
            for col in  strVec: print col
     col = evt.getCollection("Tracks")
     ntracks = col.getNumberOfElements()
     Bfield=5.0 # B-field of sidloi3 detector 
     particles=[]
     for i in range(ntracks):
          track=col.getElementAt(i)
          trk=BaseTrackState()
          trk.setZ0(track.getZ0())
          trk.setPhi(track.getPhi())
          trk.setOmega(track.getOmega())
          trk.setD0(track.getD0())
          trk.setTanLambda(track.getTanLambda())
          charge=1
          if (track.getOmega()<0): charge=-1
          mom=trk.computeMomentum(Bfield) # B filed in Z 
          px,py,pz=mom[0],mom[1],mom[2]
          ee=sqrt(px*px+py*py+pz*pz)
          p=LParticle("track",px,py,pz,ee,0)
          print(p)
     del col,evt
  reader.close() # close the file
  del reader
\end{lstlisting}

\newpage
\subsection{Running over calorimeter clusters and constructing anti-KT jets}
\label{app3}

This example deals with the construction of anti-$k_T$ jets from calorimeter clusters using 
particle-flow  algorithm objects (PFO). We read a list of files from the directory ``files'', access 
the container called "PandoraPFOCollection", 
fill the list of PFO objects and reconstruct the anti-$k_T$
jets \cite{Cacciari:2008gp}. 
The LCIO input files used by this example can be downloaded from the HepSim repository \cite{Chekanov:2014fga}.
The Java implementation
of this jet algorithm is described in \cite{Chekanov:2015cca}.
Then we print the jet collections  and fill a histogram with the jet transverse momentum. 
The canvas is saved to an image file. 

\vspace{0.5cm}
\begin{lstlisting}
from java.util import ArrayList
from hep.lcio.implementation.io import LCFactory
from jhplot import H1D
from hephysics.jet import ParticleD
from hephysics.jet import JetN2

import glob # make list of files.. 
files=glob.glob("files/*.slcio")
factory = LCFactory.getInstance()

h1=H1D("jet pt",100,0,20)
ktjet=JetN2(0.5,"antikt",20)    # antiKT with R=0.5, E-mode, anti-KT,min pT=20
print ktjet.info()              # print its settings
nEvent=0
for f in files:
  print "Open file=",f
  reader = factory.createLCReader()
  reader.open(f)
  while(1):
     evt=reader.readNextEvent()
     if (evt == None): break
     nEvent=nEvent+1
     if (nEvent%50==0): print "# Event: ",nEvent
     strVec = evt.getCollectionNames()
     if nEvent == 1:
            for col in  strVec: print col
     col = evt.getCollection("PandoraPFOCollection")
     nPFA = col.getNumberOfElements()
     alljets=[] # make a new list with jets 
     particles=ArrayList() # list of particles  
     for i in range(nPFA):
          pa=col.getElementAt(i)
          p4=pa.getMomentum()
          ee=pa.getEnergy()
          p=ParticleD(p4[0],p4[1],p4[2],ee);
          particles.add(p) # add particle to the list 
     ktjet.buildJets(particles)
     jets=ktjet.getJetsSorted()      # get a list with sorted jets
     if (len(jets)>0):
                 print "pT of a leading jet =",jets[0].perp()," GeV"
                 h1.fill(jets[0].perp())
     del col,evt
  reader.close() # close the file
  del reader

c1=HPlot("pT")
c1.visible()
c1.setAutoRange()
c1.setMarginLeft(90)
c1.setNameX("pT(jet)")
c1.setNameY("Events")
c1.draw(h1)
c1.export("mc_jets_antiKT.pdf")
\end{lstlisting}